\documentclass[10pt,a4paper,twocolumn,prb,aps,showpacs,floatfix,groupedaddress,superscriptaddress]{revtex4-2}
\usepackage{graphicx,epsfig}
\usepackage{times}
\usepackage{graphics,dcolumn,bm}
\usepackage{amssymb,amsmath,rotate,color}
\usepackage{dsfont}
\usepackage{soul}

\begin{document}
\newcommand{\be}{\begin{equation}}
\newcommand{\ee}{\end{equation}}
\newcommand{\bea}{\begin{eqnarray}}
\newcommand{\eea}{\end{eqnarray}}
\newcommand{\bseq}{\begin{subequations}}
\newcommand{\eseq}{\end{subequations}}
\newcommand{\nn}{\nonumber}
\newcommand{\vpdag}{{\vphantom{\dagger}}}
\newcommand{\bs}{\boldsymbol}
\newcommand{\up}{\uparrow}
\newcommand{\down}{\downarrow}
\newcommand{\bcpo}{BiCu$_2$PO$_6${} }
\newcommand{\bra}[1]{\langle #1|}
\newcommand{\ket}[1]{| #1\rangle}

\definecolor{red}{rgb}{1.0,0.0,0.0}
\definecolor{green}{rgb}{0.0,1.0,0.0}
\definecolor{blue}{rgb}{0.0,0.0,1.0}

\title{Interacting triplons in frustrated spin ladders:
Binding and decay in BiCu$_2$PO$_6$}

\author{Leanna B.\ M\"uller}
\affiliation{Condensed Matter Theory, 
TU Dortmund University,
Otto-Hahn-Stra\ss{}e 4, 44227 Dortmund, Germany}
\author{G\"otz S. Uhrig}
\email{goetz.uhrig@tu-dortmund.de}
\affiliation{Condensed Matter Theory, TU Dortmund University,
Otto-Hahn-Stra\ss{}e 4, 44227 Dortmund, Germany}

\date{\today}%

\begin{abstract}
Establishing a comprehensive model of the rich spin dynamics in \bcpo has been a challenge over the
last decade. Inelastic neutron scattering experiments revealed that its elementary triplons are
non-degenerate showing the existence of significant anisotropic spin couplings. Evidence for triplon decay
into two triplons has been found, but two prominent downturns in the dispersions eluded an
explanation. Level repulsion due to hybridization of single triplons with the continuum
of two-triplon scattering states has been proposed as explanation. We show that this concept
may explain the weak downturn at higher energies, but fails for the most pronounced 
downturn at lower energy. In turn, we provide evidence that this downturn is the signature 
of a two-triplon bound state of essentially singlet character
pointing to triplon-triplon interaction as the second crucial ingredient of the spin dynamics 
in this exemplary system.
\end{abstract}

\maketitle

% Introduction: General, binding, quasiparticle decay, avoided decay
\section{Introduction} 

Strongly correlated quantum systems are notoriously difficult to understand due to the
extremely fast growth of dimensionality of the relevant Hilbert space with system size. 
At low energies close to the ground state, however, the situation is much more
favorable. Elementary excitations can be identified in terms of which most
physical properties can be explained; they can be addressed as particles or
quasiparticles of the system. For translationally invariant systems, i.e., in 
crystals, the dispersion $\hbar\omega(\vec k)$ yields the energy dependence on 
the momentum $\hbar \vec k$ of the quasiparticles. 
Theoretically, this is described by a diagonal Hamiltonian which is bilinear
in creation and annihilation operators.
In addition, these excitations can decay, for instance one excitation into two
excitations which is captured by trilinear terms in the Hamiltonian. The pairwise
interaction of two particles, called two-particle interaction, is described by 
quadrilinear terms.
Terms and processes involving even more quasiparticles may also arise and 
can be denoted by quintilinear terms and sextilinear terms (three-particle interactions) 
and so on. These higher interactions are rarely considered, but 
they do occur in effective models and can lead to important shifts of 
spectral weight and even bound states \cite{schmi22}.

Here we focus on systems formed by localized spins in insulators which are called quantum magnets.
Generically, they provide clean and well-defined systems which are only weakly coupled
to other degrees of freedom such as charges or phonons. Thus, the elementary excitations
are long-lived unless they decay intrinsically due to terms beyond the bilinear ones.
Omnipresent pairwise interactions can lead to bound states. Hence, quasiparticle decay
and binding are the two foci of the present Letter which we discuss in the particular
system \bcpo realizing weakly coupled spin ladders \cite{tsirl10}.

There are three large classes of elementary excitations of quantum magnets. In 
systems with long-range magnetic order the elementary excitations are 
quantized spin waves, so called magnons. In low-dimensional or very strongly correlated
systems fractional excitations occur called spinons. In gapped quantum antiferromagnets 
without order the elementary excitations above the $S=0$ ground state have triplet
character with $S=1$ and are called triplons \cite{schmi03c}
to underline their quasiparticle nature.
For \bcpo, these triplons are the relevant elementary excitations 
\cite{plumb13,plumb16,splin16,hwang16}

Theoretically, quasiparticle decay and its preconditions have been discussed 
generally \cite{pitae59,gavea95} and for magnetic excitations in particular 
\cite{zhito06,fisch10a,fisch11b,zhito13,verre19}. Experimentally, 
it has been observed in a number of ordered quantum magnets \cite{hong17a,oh13,robin14}
and disordered quantum magnets \cite{stone06,masud06,plumb16} 
underlining the relevance of the issue. The decay of magnons is particularly strong
if a single magnon can decay into two magnons as is the case for non-collinear
ordering, for instance in triangular quantum antiferromagnets 
\cite{zheng06b,chern09,oh13,mouri13b,mouri13b_err,verre19}. A noteworthy observation
is that the trilinear terms also imply a renormalization of the energy
of the single magnon by level repulsion. For large enough trilinear terms
the dispersion is pushed below the multiparticle continuum so that
no decay occurs anymore; this has been dubbed ``avoided quasiparticle decay'' \cite{verre19}. 
Generically, it is also seen in one-dimensional systems \cite{fisch10a,fisch11b,verre19}.
Similarly, decay processes from one quasiparticle to three quasiparticles
also lead to a renormalization of the dispersion due to level repulsion as 
seen in the collinear square lattice Heisenberg antiferromagnet \cite{powal15,powal18,verre18b}.

% noch was allg. zu Bindungen
Binding stems from the quadrilinear terms as explained above. In quantum ferromagnets
the fully polarized state is the ground state and the number of magnons is conserved
in the model as given. Since long, bound states of magnons in ferromagnets
are established \cite{bethe31,dyson56a,hanus63,worti63} and 
play a role in current applications \cite{barke13}.
For quantum antiferromagnets, the situation has more facets depending 
on the nature of the ground state. For unfrustrated antiferromagnets with
long-range order, bound states of two magnons appear in gapped, easy-axis
systems \cite{oguch73,hamer09,dusue10,verre18b} which disappear on passing to
the isotropic antiferromagnet. But there remains a strong attractive force
which shifts spectral weight to lower energies and induces a roton-like
minimum in the magnon dispersion by level repulsion \cite{powal15,powal18}.

%Spinonen
In one dimension, no long range order occurs without
spin anisotropies in the Hamiltonian. 
But fractional spinon excitations are established \cite{fadde81}
which also form bound states \cite{bethe31,kohno09,ganah12}. They could 
be experimentally verified only recently \cite{wang18a}. We point out that triplon excitation
can be also seen as bound states of confined spinons \cite{greit02a,kohno07}.

Gapped antiferromagnets with neither long-range order 
nor fractionalization are valence bond solids
and their elementary excitations are triplons which
generically attract each other, in particular if the total spin remains zero.
This has been established for spin ladders 
\cite{shelt96a,sushk98,trebs00,knett01b,windt01,schmi05b,notbo07},
dimerized spin chains \cite{tsvel92,uhrig96b,uhrig96be,schmi04a}, 
and for the Shastry-Sutherland lattice \cite{knett00b,lemme00a,knett04a}, 
both in theory and experiment.

% Dann BCPO
\section{The Compound}

\begin{figure}[htb]
\includegraphics[width=4.5cm]{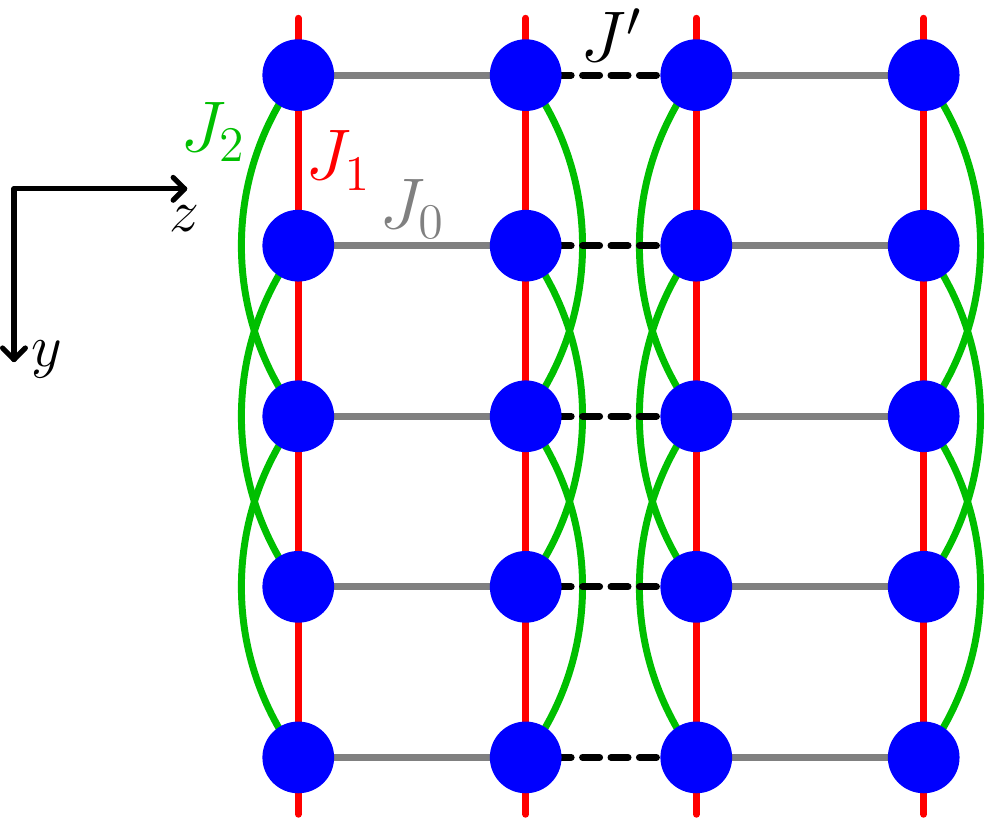}
\hspace*{3.8cm}

\vspace*{-30mm}

\hspace*{4.9cm}
\begin{tabular}[b]{c|c|c}
$D_{ij}^{\alpha}$ & along legs & parity\\
\hline
$D_{0}^{y}$ & alternating & odd\\
$D_{1}^{x}$ & uniform & odd\\
$D_{1}^{y}$ & alternating & odd\\
$D_{2}^{x}$ & uniform & odd\\
$D_{2}^{z}$ & alternating & even
\end{tabular}
\vspace*{7mm}
\caption{Left: \textbf{Sketch of the spin model for \bcpo}; blue spheres denote the Cu$^{2+}$ ions hosting the 
spins $S=1/2$. The isotropic  couplings $J_0$, $J_1$ and $J_2$ 
define a single frustrated ladder while the isotropic interladder coupling 
$J^{\prime}$ weakly links adjacent ladders forming planes. Right:
Sign of non-vanishing DM terms along the legs, parity w.r.t.\ reflection.  
The parity given for $D_{0}^{y}$ refers to its term in the Hamiltonian, not to $D_{0}^{y}$
itself.}
\label{fig:mag_model}
\end{figure}

In this Letter, we elucidate quasiparticle decay and binding in the strongly frustrated
spin ladders in \bcpo. The crystal structure has been discussed by Tsirlin \textit{et al.}
\cite{tsirl10}, but for our purposes the spin model is sufficient. The spins $S=1/2$
are localized at the copper ions which are positioned in tubes made from an upper
and a lower spin laddder. In addition, the spin ladders are weakly coupled
forming a two-dimensional lattice \cite{mentr06,wang10c}. Here only
the topology of the magnetic exchange couplings matters so that we project the 
upper and lower spin ladders in one plane, see Fig.\ \ref{fig:mag_model}.

The isotropic part of the Hamiltonian of a single frustrated spin ladder reads 
\begin{equation}
\label{isotropic_spin_ladder}
\mathcal{H}_{\mathrm{lad}}=J_0\sum\limits_{i}^{}\mathbf{S}_{i}^{\mathrm{L}}
\mathbf{S}_i^{\mathrm{R}}+J_1\sum\limits_{i,\tau}^{}
\mathbf{S}_{i}^{\tau}\mathbf{S}_{i+1}^{\tau}+
J_2\sum\limits_{i,\tau}^{}\mathbf{S}_{i}^{\tau}\mathbf{S}_{i+2}^{\tau}
\end{equation}
with the rung index $i$ and the $\tau\in\{\text{L, R}\}$ indicating the left or right leg, respectively. 
The interladder coupling $J'$ is significantly weaker than the intraladder couplings, but
not negligible \cite{plumb13,tsirl10}. 
\bcpo represents a valence bond solid with triplons as elementary magnetic excitations
because its spin system is gapped.
Since the inversion symmetry about the Cu-Cu bonds is broken, anisotropic 
Dzyaloshinskii-Moriya (DM) couplings may occur \cite{moriy60b}
and were conjectured \cite{tsirl10}. The splitting of the otherwise degenerate triplons 
\cite{plumb13,plumb16} clearly confirm this conjecture.

Including anisotropic couplings to our model of \bcpo we discuss the Hamiltonian
\begin{equation}
\label{complete_hamiltonian}
\mathcal{H}=\mathcal{H}_{\mathrm{lad}}+\sum\limits_{i,j}^{}\mathbf{D}_{ij}\left(\mathbf{S}_{i}\times\mathbf{S}_j\right)
+\sum\limits_{i,j}^{}\sum\limits_{\alpha,\beta}^{}\Gamma_{ij}^{\alpha\beta}S_{i}^{\alpha}S_{j}^{\beta}.
\end{equation}
Note that we include the label $\{ \text{L}, \text{R}\}$ indicating the
leg of the ladder in the site indices $i$ and $j$ henceforth.
In addition to $\mathcal{H}_{\mathrm{lad}}$ from Eq.\ \eqref{isotropic_spin_ladder}
we consider the DM interactions $\mathbf{D}_{ij}\left(\mathbf{S}_{i}\times\mathbf{S}_j\right)$
and the symmetric anisotropic exchanges $\Gamma_{ij}^{\alpha\beta}S_{i}^{\alpha}S_{j}^{\beta}$. 
The sums with the site indices $i$ and $j$ are meant to count each pair of spins once. 
The couplings along the rungs of the spin ladder are labeled with the index $0$, 
thus $J_0$, $\mathbf{D}_0$ and $\Gamma_0^{\alpha\beta}$. The index $1$ is used for all couplings between other nearest-neighbor
(NN) sites, thus  $J_1$, $\mathbf{D}_1$ and $\Gamma_1^{\alpha\beta}$. 
Lastly, the couplings between next-nearest neighbor (NNN) sites are marked with the index $2$, 
thus $J_2$, $\mathbf{D}_2$ and $\Gamma_2^{\alpha\beta}$; see also Fig. \ref{fig:mag_model}. 
The symmetric anisotropic exchanges $\Gamma_{ij}^{\alpha\beta}$ represent the second order
effecs of the DM terms \cite{shekh92} reading 
\begin{equation}
\label{formular_gamma}
\Gamma_{ij}^{\alpha}=\frac{D_{ij}^{\alpha}D_{ij}^{\beta}}{2J_{ij}}
-\frac{\delta^{\alpha\beta}\mathbf{D}_{ij}}{6J_{ij}}
\end{equation}
 and being defined such that the tensor $\Gamma_{ij}$ is traceless \cite{splin16}. 
The orientation of the $\mathbf{D}$ vectors is determined by the selection rules based on the point group symmetries
\cite{moriy60b}, see Fig. \ref{fig:mag_model}. The symmetry analysis \cite{splin16} yields the results 
displayed on the r.h.s.\ of Fig.\ \ref{fig:mag_model}. Five DM components can assume finite values from which
four components hold odd parity with respect to reflection about the center line of the spin ladder \cite{schmi05b}.
Hence they do not contribute to bilinear terms in the Hamiltonian, but only to trilinear and potential
quintilinear terms.

\section{Theoretical Analyses}

The approach used extends the treatment in Ref. \onlinecite{splin16}.
We start from the isolated isotropic spin ladder which we map by continuous
unitary transformation (CUT) \cite{knett00a,krull12} to an effective model formulated
in second quantization, i.e., in creation and annihilation operators of the elementary 
excitations, triplons \cite{schmi03c,kohno07,sachd08}. 
The CUT takes the hardcore property of the triplons fully into account.
This effective model conserves the number
of triplons so that the computation of physical properties such as the dispersion,
bound states, or spectral densities is greatly facilitated. 

The CUT represents a 
systematic basis change
and can be applied as well to any observable, in particular to the spin components $S_{i,\alpha}$. 
These transformed observables allow us to express all 
residual couplings (antisymmetric ($\propto{\bf D}_{ij}$) and symmetric DM terms 
($\propto\Gamma^{\alpha\beta}_{i,j}$),
interladder couplings ($\propto J'$)) in terms for triplon operators. 
Since these additional couplings are small we ignore the hardcore property of the
triplon operators in the subsequent step so that we can diagonalize the bilinear
parts by a Bogoliubov transformation. This transformation is also applied to 
the trilinear terms ensuing from the odd parity DM terms. The quadrilinear terms are 
taken from the CUT of the isolated ladders and also subjected to the Bogoliubov transformation.
In view of the three momenta which can be varied freely in 
quadrilinear terms with momentum conservation this comprises a particularly large
number of coefficients representing the numerical bottleneck in this calculation.
From the resulting terms we keep the two-triplon interaction terms consisting
of monomials of two creation and two annihilation operators. They are responsible for
binding phenomena.

Finally, we compute resolvents in the hybridized one-triplon and two-triplon subspace
at given total momentum $\hbar{\bf q}$. These resolvents provide access to the dynamic structure 
factor $S({\bf q},\omega)$ including sharp resonances outside the continuum of
scattering states, see Appendix \ref{app:details}. The momentum dependence of these sharp resonances is to be compared
with the dispersion of the experimentally observed strong peaks.
Sharp resonances can stem from either single triplon states which are renormalized
by the hybridization with the two-triplon states or from bound two-triplon states.

\paragraph{\bf Bilinear Level}
Previously, we performed the analysis of experimental inelastic neutron scattering (INS) data for \bcpo
on the bilinear level and achieved best fits with the parameters $J_0=9.4$meV, $J_1=1.2J_0$, 
$J_2=0.9J_1$, $J'=1.5$meV, $D_0^y=0$, $D_1^x=0.48 J_1$, $D_1^y=0.61 J_1$, $D_2^x=0$, and 
$D_2^z=-0.02 J_2$. A figure showing the 
degree of agreement achieved is included in Appendix \ref{app:bilinear}; for details
see Ref.\ \cite{splin16}. Summarizing, the dispersions in the vicinity of the minima
is captured well. But the downturns of the measured data remain completely unexplained.
Additionally, the values of the NN DM terms are substantially larger than one is expecting
for the relative strength of couplings stemming from spin-orbit coupling: 10 to 20\%
of the corresponding exchange coupling. Qualitatively, these findings agree with those 
by Hwang and Kim \cite{hwang16} for the bilinear model.

\begin{figure}[htb]
\includegraphics[width=\columnwidth]{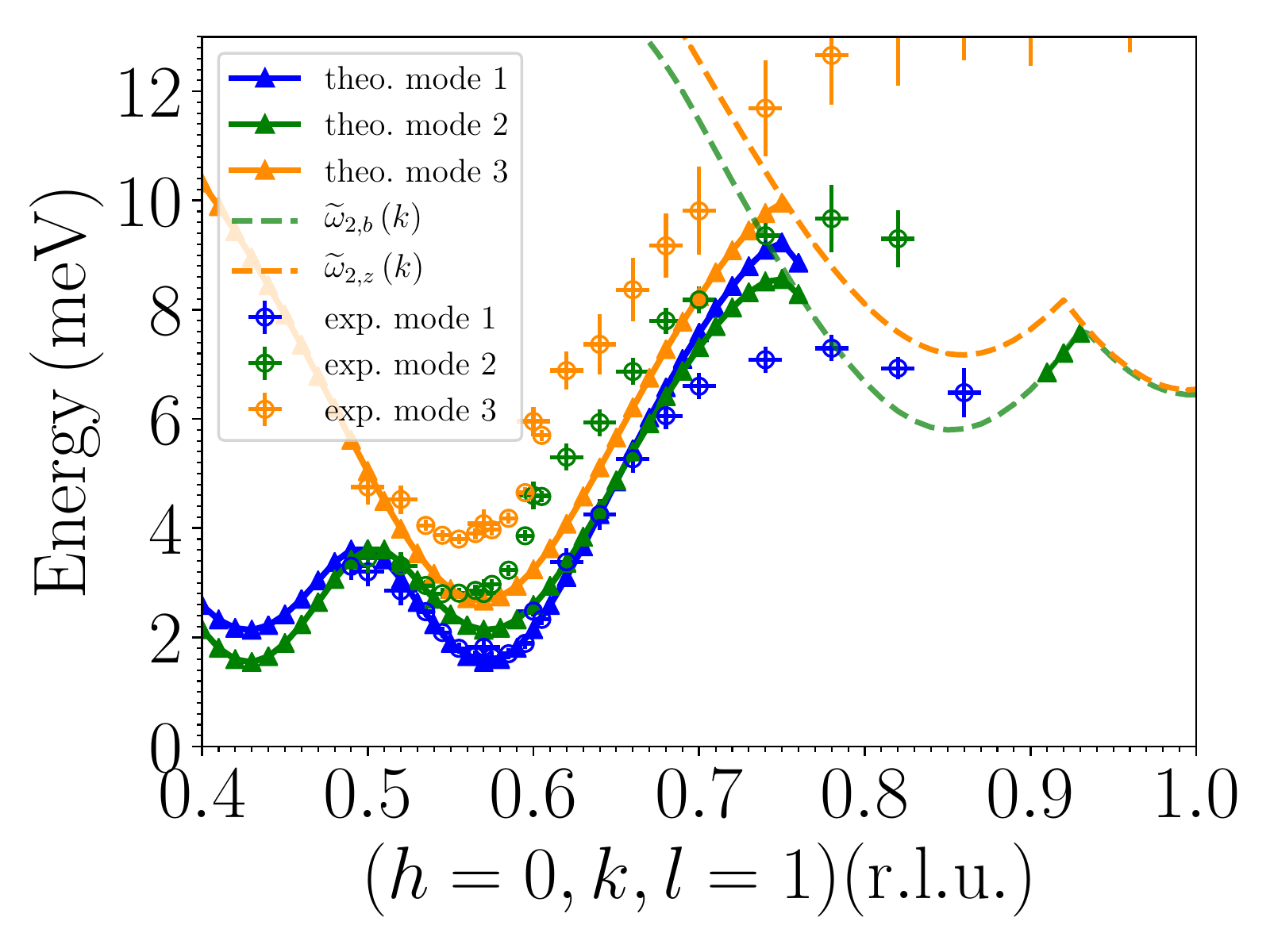}
\caption{\textbf{Dispersion in the trilinear model.}
Energetically low-lying dispersion including the hybridization between
the one- and the two-triplon states. A weak down-bending occurs before the
single triplon dispersion enters the two-triplon continua at about $k=0.75$r.l.u..
The parameters used are $J_0=8.0$meV, $J_1=1.2J_0$, $J_2=0.9J_1$,
$J'=0.16J_0$, $D_0^y=0.1J_0$, $D_1^x=0.25 J_1$, $D_1^y=0.38 J_1$, $D_2^x=0.0$, and 
$D_2^z=-0.09 J_2$.  The dashed lines represent the lower edges of the two-triplon continua.
The yellow curve is computed from the highest, yellow dispersion. The green one from the 
green dispersion. All other  four continua edges lie between the two depicted continuum edges.
Unlinked symbols are the experimental data from Ref.\ \cite{plumb14,plumb16}.}
\label{fig:trilinear}
\end{figure}

%%% trilinear
\paragraph{\bf Trilinear Level}
This brings us to the trilinear terms resulting from the
DM couplings of odd parity. As explained above, 
the trilinear terms introduce a new piece of physics,
namely the decay of a triplon into two triplons and vice-versa
the fusion of two triplons to a single one. 
In short, the one- and the two-triplon subspaces hybridize. 
The results are shown in Fig.\ \ref{fig:trilinear}. The dispersion
branches around the minima are captured, but less well than
on the bilinear level. This is due to the much more 
demanding numerical evaluation which does not allow us to
scan all conceivable parameter combinations, for instance
we stick to the ratios $J_1/J_0$ and $J_2/J_1$ determined on the bilinear
level.

One  satisfying observation underlining the importance of the
trilinear terms is that the required DM couplings are much lower than
on the bilinear level. The value for $D_1^x$ has lowered from $0.48J_1$ to 
$0.25 J_1$ and for $D_1^y$ from $0.61J_1$ to  $0.38 J_1$ which brings them
closer to the expected range of spin-orbit induced couplings.
On the downside, however, we only find a weak feature of level repulsion
in Fig.\ \ref{fig:trilinear}. There is a small downturn in the
dispersion just before it enters into the scattering continuum
at $k\approx 0.75$ r.l.u.. Comparing this feature to the significant experimental 
downturn at 7 meV it is evident that the hybridization due to
trilinear terms is insufficient to explain the measured features. 
We stress that this conclusion agrees
with the figures from previous calculations \cite{hwang16}.
We deduce that the level repulsion from the scattering continuum alone
is too weak to account for the observed dispersions.
This finding raises the question what generates
 the experimental features. Promising candidates are bound states
since it is known that spin ladders host bound two-triplon states
\cite{shelt96a,sushk98,trebs00,knett01b,windt01,schmi05b,notbo07}. 
In addition, frustration enhances binding because the mobility
of triplons is reduced while its interaction is enhanced 
\cite{uhrig96b,uhrig96be,schmi04a}.

\begin{figure}[htb]
\includegraphics[width=\columnwidth]{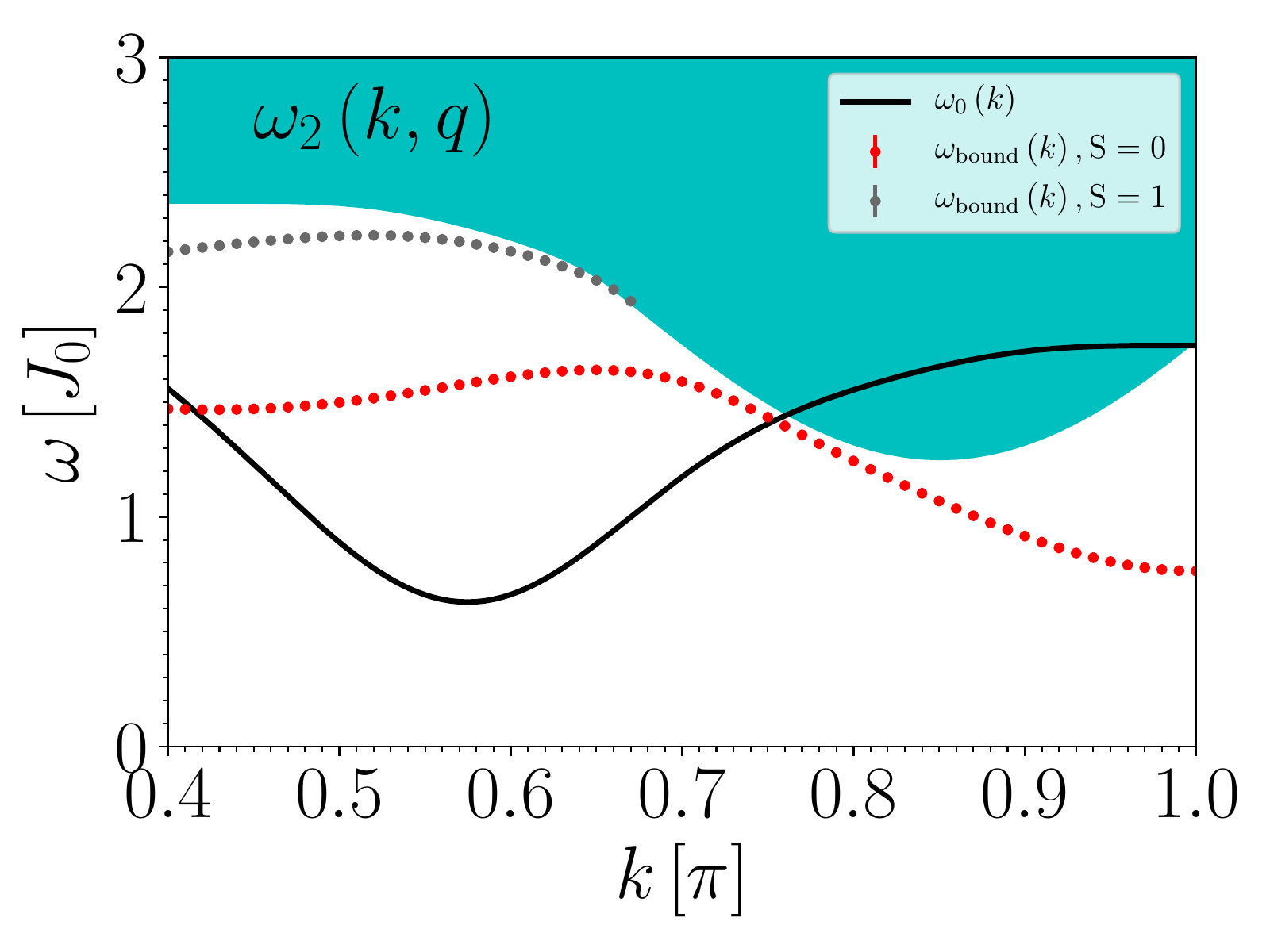}
\caption{\textbf{Isolated spin ladder} for $J_1=1.2J_0$
and $J_2=0.9J_1$: one-triplon dispersion in black, two-triplon scattering
continuum is colored; the $S_\text{tot}=0$ bound state is shown by
the dotted red line and the $S_\text{tot}=1$ bound state by the dotted grey line.
Note that the wave number is given in units of $\pi$ setting the lattice
constant to unity. This corresponds to the reciprocal lattice units of \bcpo where
the actual lattice constant is the distance between next-nearest neigbor rungs, cf.\
Refs.\ \cite{tsirl10,splin16}.}
\label{fig:binding-isolated}
\end{figure}

\paragraph{\bf Quadrilinear Level}
Describing the binding of two particles requires 
to include the quadrilinear terms in second quantization. First, we inspect the isolated ladder 
for which the CUT alone yields the effective model very reliably. 
Fig.\ \ref{fig:binding-isolated}
displays the dispersions of a single triplon and of the two bound
states formed by two triplons. The latter are situated below the two-triplon
band of scattering states shaded in light green-blue.
It is remarkable that just at $k\approx 0.75$ r.l.u. the $S_\text{tot}=0$ state,
displayed by the red dotted curve, falls below the single triplon dispersion. 
This suggests that
a bound state could indeed be involved in the experimental situation.
A counterargument is that INS is sensitive only to $S=1$ states,
not to $S=0$ states. While this holds in spin isotropic
systems it does not apply in presence of the sizable spin anisotropies 
existing in \bcpo. For completeness, we also depict the bound $S=1$ state 
by the gray dotted curve.

\begin{figure}[htb]
\includegraphics[width=\columnwidth]{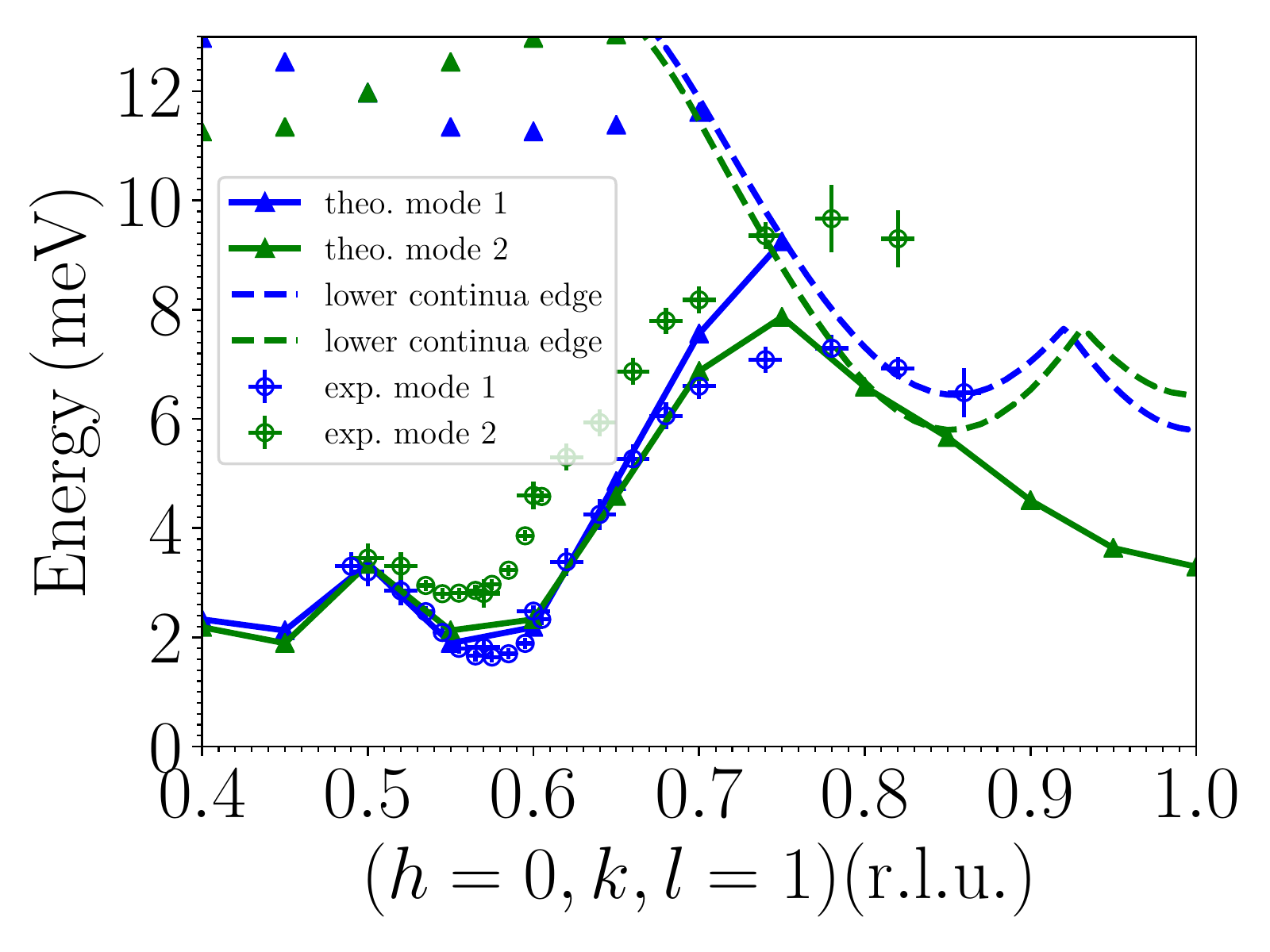}
\caption{\textbf{Energies in the quadrilinear model} for \bcpo for the parameters
$J_0=8.0$meV, $J_1=1.2J_0$, $J_2=0.9J_1$,
$J'=0.16J_0$, $D_0^y=0.1J_0$, $D_1^x=0.25 J_1$, $D_1^y=0.38 J_1$, $D_2^x=0.0$, and 
$D_2^z=-0.06 J_2$. The green (blue) dashed line shows the lower continuum edge derived
from the theoretical dispersion displayed as green (blue) solid line.
Unlinked symbols are the experimental data from Ref.\ \cite{plumb14,plumb16}.}
\label{fig:quadrilinear}
\end{figure}

% passing on to the full model: Discussion
As next step, we compute the dynamic structure factor including the triplon-triplon 
interaction in the full 2D model in reciprocal space with in the Hilbert space
spanned by one-triplon and two-triplon states at given total momentum. 
This is a very demanding calculation because the matrix representing the Hamiltonian
is no longer sparse. In the previous trilinear case, the Hamiltonian relevant for
the two-triplon subspace is diagonal. The triplon-triplon interaction, however, 
implies scattering between the states of all relative momenta so that the 
corresponding matrix is dense. We performed calculations for $40\times 8$ 
${\bf k}$-points. The results are shown in Fig.\ \ref{fig:quadrilinear}.
Clearly, an additional branch with significant weight forks off the 
dispersion of the single dispersion at $k\approx 0.75$ r.l.u. with $\approx 7$\,meV. This agrees 
very well with the most prominent downturn in experiment. It provides evidence
that two-triplon interactions and the concomitant binding are essential players in the
physics of \bcpo. The bound state is not observed in experiment
to $1\text{r.l.u.}$ because it is an $S_\text{tot}=0$ and visible only
due to the DM terms. Beyond $0.8\text{r.l.u.}$ its weight becomes 
very small \cite{mulle21}.

Unfortunately, the resolution of the present calculation is not high enough
to make a tangible statement about the second downturn at higher energy around $9.5$\,meV.
It could result from a weaker bound state formed from triplons of different flavors.
Note that the spin anisotropies split the triplons and hybridize them so that
they acquire mixed flavors. An alternative and appealing explanation for the weak downturn
at higher energies is suggested by the results for the trilinear model, see Fig.\ 
\ref{fig:trilinear}. While they do not explain the strong downturn at about 7\,meV, 
it is possible that they explain the less pronounced downturn around $9.5$\,meV
due to level repulsion \cite{plumb16}.
The wave vector and the energy at which this downturn occurs agree between theory and
experiment as well as the overall size of the feature. These observations call for
future high-precision calculations including trilinear \emph{and} quadrilinear terms.
In this way, the intricate interplay between hybridization and attractive interaction
in \bcpo can be elucidated quantitatively. The lesson learned will provide the foundation
for understanding many more strongly frustrated low-dimensional quantum antiferromagnets
where tri- and quadrilinear processes need to be measured and understood.

\section{Conclusion}

Summarizing, we analyzed the unusual dispersions found in \bcpo with respect to
two important phenomena: decay and binding. Decay results from trilinear
terms in second quantization formed by two creation and one annihilation operator
or their hermitian conjugate representing fusion. Binding stems from
quadrilinear terms formed by two creation and two annihilation operators.
Generically, both types are present and their importance relative to the single-particle
dispersion is particularly high in low-dimensional systems. \bcpo is
a challenging exemplary system to study these fundamental processes because
it hosts significant spin anisotropy terms so that different triplon flavors
can be distinguished by their energy. In this system, evidence
for the relevance of trilinear terms was found \cite{plumb16,hwang16}. Yet
the strong, dominant downturn eluded an explanation. Our analysis including
triplon-triplon interaction, i.e., with quadrilinear terms, 
suggests that the strong downturn is the signature of a $S=0$ two-triplon
bound state, slightly perturbed by the anisotropic terms so that it
becomes visible in the dynamic structure factor measured by inelastic neutron
scattering. The weak downturn at higher energies can be identified as the signature of
level repulsion between a single triplon and the two-triplon continuum.
These findings suggest \bcpo as a challenging exemplary platform to investigate
the phenomena of binding and decay of multi-flavor quasi-particles.

\begin{acknowledgments}
This study was funded by the German Research Foundation (DFG) in
grant UH 90/14-1. Parts of the calculations were performed on the LiDO3
high-performance computing system funded by the
DFG. We acknowledge useful discussions with Nils Drescher;
he also provided the code to perform the CUT. We thank Carsten Nase for
lasting technical support.
\end{acknowledgments}

%\bibliography{../../AAA_bibinput/liter10}


\begin{appendix}

\section{Technical Details}

\label{app:details}

The dynamic structure factor in the time domain reads
\begin{equation}
S^{\alpha\beta}(\mathbf{q},\omega)=\frac{1}{2\pi}
\int\limits_{-\infty}^{\infty}\mathrm{d}t\,
\mathrm{e}^{\mathrm{i}\omega t}\langle S^{\alpha}(-\mathbf{q},t)S^{\beta}(\mathbf{q},0)\rangle.
\end{equation}
At zero temperature, its Fourier transform for $\alpha=\beta$
is given by the retarded Green function
\begin{equation}
\label{eq:green}
G^{\mathrm{ret},\alpha}\left(\mathbf{q},\omega\right)=\bra{u_0}\frac{1}{\omega-{H(\mathbf{q})}}\ket{u_0},
\end{equation}
where $\ket{u_0}=S^{\alpha}(\mathbf{q},0)|0\rangle$ with $|0\rangle$ being the ground state. 
We omit $\alpha$ henceforth since we focus on the energies
and not on the spectral weights here.

The action of the Hamiltonian in  \eqref{eq:green} is represented by a tridiagonal matrix
\begin{equation}
\bar{H}\left(\mathbf{q}\right)=
\begin{pmatrix}
a_{0}\left(\mathbf{q}\right) & b_{1}\left(\mathbf{q}\right) & 0 & 0 & \cdots\\
b_{1}\left(\mathbf{q}\right) & a_{1}\left(\mathbf{q}\right) & b_{2}\left(\mathbf{q}\right) & 0 & \cdots\\
0 & b_{2}\left(\mathbf{q}\right) & a_{3}\left(\mathbf{q}\right) & b_{3}\left(\mathbf{q}\right) & \cdots\\
\vdots & \vdots & \vdots & \vdots & \ddots ,
\end{pmatrix}
\end{equation}
which is computed by Lanczos iteration
\begin{subequations}
\begin{alignat}{2}
&\ket{u_{1}}&&=\left(H\left(\mathbf{q}\right)-a_{0}\left(\mathbf{q}\right)\right)\ket{u_0}\\
&\ket{u_{2}}&&=\left(H\left(\mathbf{q}\right)-a_{1}\left(\mathbf{q}\right)\right)\ket{u_1}-b_1^2\left(\mathbf{q}\right)\ket{u_0}\\
&\ket{u_{3}}&&=\left(H\left(\mathbf{q}\right)-a_{2}\left(\mathbf{q}\right)\right)\ket{u_2}-b_2^2\left(\mathbf{q}\right)\ket{u_1}\\
&...&&\phantom{=} \nonumber
\end{alignat}
\end{subequations}
yielding the coefficients
\begin{subequations}
\begin{alignat}{2}
\label{formular_lanczos_coefficient_a}
a_{i}\left(\mathbf{q}\right)&=\frac{\bra{u_{i}}H\left(\mathbf{q}\right)\ket{u_i}}{\langle u_i|u_i\rangle}\quad&&\text{for\,\,}i=0,1,2,...\\
b_{i}^2\left(\mathbf{q}\right)&=\frac{\langle u_i|u_i\rangle}{\langle u_{i-1}|u_{i-1}\rangle}\quad&&\text{for\,\,}i=1,2,3,...\\
b_{0}\left(\mathbf{q}\right)&=0.\quad&&\phantom{=}
\end{alignat}
\end{subequations}

Finally, the Green function is computed by its continued fraction down to a certain depth
of 30 to 100 fractions
\begin{equation}
G^{\mathrm{ret}}\left(\mathbf{q},\omega\right)=\frac{1}{\omega-a_{0}\left(\mathbf{q}\right)-
\frac{b_1^2\left(\mathbf{q}\right)}{\omega-a_{1}\left(\mathbf{q}\right)-\frac{b_2^2\left(\mathbf{q}\right)}{\omega-a_{2}\left(\mathbf{q}\right)-\cdots}}},
\end{equation}
Then, the continued fraction is terminated by the square root terminator
\begin{subequations}
\begin{align}
T\left(\mathbf{q},\omega\right)&=\frac{1}{2b_{\infty}^2\left(\mathbf{q}\right)}\left(\omega - a_{\infty}\left(\mathbf{q}\right)-\sqrt{R\left(\mathbf{q},\omega\right)}\right)\nonumber\\
&\text{for\,\,}\omega\geq\omega_{2,\mathrm{max}}\left(\mathbf{q}\right)\\
T\left(\mathbf{q},\omega\right)&=\frac{1}{2b_{\infty}^2\left(\mathbf{q}\right)}\left(\omega - a_{\infty}\left(\mathbf{q}\right)-\mathrm{i}\sqrt{-R\left(\mathbf{q},\omega\right)}\right)\nonumber\\
&\text{for\,\,}\omega_{2,\mathrm{min}}\leq\omega\leq\omega_{2,\mathrm{max}}\left(\mathbf{q}\right)\\
T\left(\mathbf{q},\omega\right)&=\frac{1}{2b_{\infty}^2\left(\mathbf{q}\right)}\left(\omega - a_{\infty}\left(\mathbf{q}\right)+\sqrt{R\left(\mathbf{q},\omega\right)}\right)\nonumber\\
&\text{for\,\,} \omega\leq\omega_{2,\mathrm{min}}\left(\mathbf{q}\right),
\end{align}
\end{subequations}
where we used
\begin{subequations}
\label{lanczos_boarders}
\begin{align}
\omega_{2,\mathrm{min}}\left(\mathbf{q}\right)&=a_{\infty}\left(\mathbf{q}\right)-2b_{\infty}\left(\mathbf{q}\right)\\
\omega_{2,\mathrm{max}}\left(\mathbf{q}\right)&=a_{\infty}\left(\mathbf{q}\right)+2b_{\infty}\left(\mathbf{q}\right)
\end{align}
\end{subequations}
and
\begin{equation}
R\left(\mathbf{q},\omega\right)=\left(\omega - a_{\infty}\left(\mathbf{q}\right)\right)^2-4b_{\infty}^2\left(\mathbf{q}\right).
\end{equation}

\section{Dispersion in the bilinear model}

\label{app:bilinear}

\begin{figure}[htb]
\includegraphics[width=\columnwidth]{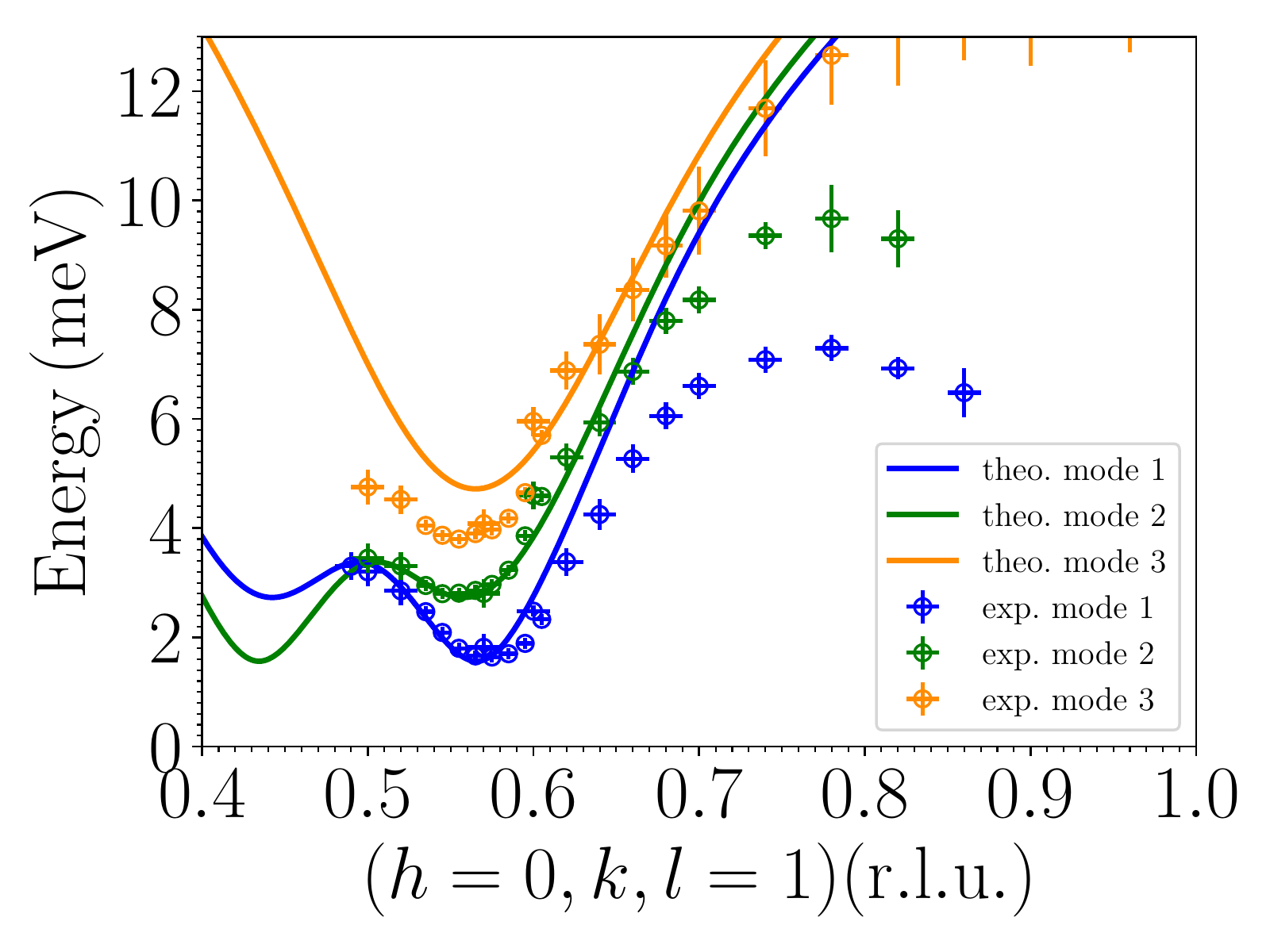}
\caption{\textbf{Energies in the bilinear model} for \bcpo for the parameters 
$J_0=9.4$meV, $J_1=1.2J_0$, $J_2=0.9J_1$, $J'=1.5$meV, $D_0^y=0$, $D_1^x=0.48 J_1$, 
$D_1^y=0.61 J_1$, $D_2^x=0$, and $D_2^z=-0.02 J_2$. 
 Unlinked symbols are the experimental data from Ref.\ \cite{plumb14,plumb16}.}
\label{fig:bilin}
\end{figure}

An overview of the dispersion in the bilinear model including the description
of all technical aspects is provided in Ref.\ \cite{splin16}. For comparison to the
dispersions on the trilinear and on the quadrilinear level we include here
the bilinear data in Fig.\ \ref{fig:bilin}. The theoretical fits agree well with
the experimental data, but the bilinear theory does not capture any feature
like the downturns. They require to consider models beyond the 
bilinear level.

\end{appendix}

\end{document}